\documentclass[aps,prb,twocolumn,groupedaddress]{revtex4}
\usepackage{graphicx}
\usepackage{amsmath,amssymb,amsfonts}
\usepackage{natbib}
\usepackage{dcolumn}
\usepackage{bm}

\newcommand{\beq}{\begin{equation}}
\newcommand{\beqa}{\begin{eqnarray}}
\newcommand{\eeq}{\end{equation}}
\newcommand{\eeqa}{\end{eqnarray}}

\begin{document}

\title{Quantum Zeno-based control mechanism for molecular
fragmentation}

\author{C. Sanz-Sanz}
\affiliation{Instituto de F\'{\i}sica Fundamental -- CSIC,
Serrano 123, 28006 Madrid, Spain.}

\author{A. S. Sanz}
\affiliation{Instituto de F\'{\i}sica Fundamental -- CSIC,
Serrano 123, 28006 Madrid, Spain.}

\author{T. Gonz\'alez-Lezana}
\affiliation{Instituto de F\'{\i}sica Fundamental -- CSIC,
Serrano 123, 28006 Madrid, Spain.}

\author{O. Roncero}
\email{octavio.roncero@csic.es}
\affiliation{Instituto de F\'{\i}sica Fundamental -- CSIC,
Serrano 123, 28006 Madrid, Spain.}

\author{S. Miret-Art\'es}
\affiliation{Instituto de F\'{\i}sica Fundamental -- CSIC,
Serrano 123, 28006 Madrid, Spain.}


\begin{abstract}
 A quantum
control mechanism is proposed for molecular fragmentation
processes within a scenario grounded on the quantum Zeno effect.
 In particular, we focus on the van der Waals Ne-Br$_2$
complex, which displays two competing dissociation channels via
vibrational and electronic predissociation. Accordingly, realistic
three dimensional wave packet simulations are carried out by using
{\it ab initio} interaction potentials recently obtained to
reproduce available experimental data. Two numerical models to
simulate the repeated measurements are reported and analyzed. It
is found that the otherwise fast vibrational predissociation is
slowed down in favor of the slow electronic (double fragmentation)
predissociation, which is enhanced by several orders of magnitude.
Based on these theoretical predictions, some hints to
experimentalists to confirm their validity are also proposed.
\end{abstract}



\maketitle


The quantum Zeno effect (QZE) was theoretically predicted in 1977
by Misra and Sudarshan\cite{misra} and experimentally observed by
Itano {\it et al.}\cite{itano} almost 15 years later. The
opposite phenomenon, the so-called anti-Zeno effect (AZE), was
later on predicted\cite{kurizki,kaulakys} and
observed immediately afterwards.\cite{fischer}
These are well-known phenomena in the literature
of quantum theory of measurement and open quantum systems, mainly
in connection with decoherence processes.\cite{breuer}
If a series of successive observations or measurements are carried out
on an unstable system at very short time intervals $\tau$, its decay
can be either inhibited (QZE) or accelerated (AZE). These two
remarkable behaviors with time become apparent when looking at the
{\it survival probability} $P(t)$, namely the probability to find
the system in the same initial state, $|\psi (0)\rangle$, at a
certain subsequent time $t$,
\begin{equation}
\label{survival}
 P(t) = |\langle \psi (0) | \psi (t) \rangle|^2
  = |\langle \psi (0) | e^{- i \hat{H} t / \hbar} | \psi (0) \rangle|^2 .
\end{equation}
Here, $\hat{H} = \hat{H}_0 + \hat{V}$ encompasses the system
free-evolution Hamiltonian, $\hat{H}_0$, and the system interaction,
$\hat{V}$.
Because of the continuous measurement (or projection) process, the
system decay can be described, in general, as\cite{kurizki}
\begin{equation}
\label{pt}
 P(t) \sim e^{- \gamma(\tau) t} ,
\end{equation}
where $\gamma(\tau)$ is the measured-modified decay rate, which depends
on $\tau$. This rate is different from the free-measurement system
decay rate, $\gamma_0$.
The decay rate $\gamma(\tau)$ can be expressed\cite{kurizki} in terms
of the spectral density of final states, $G(\omega)$, and the
measured-induced initial state level width, $F(\omega;\tau)$, as
\begin{equation} \label{gamma}
 \gamma (\tau) = 2\pi\int_0^\infty G(\omega) F(\omega;\tau) d \omega .
\end{equation}
The measurement frequency $\nu= 1/\tau$ is related to the initial
state energy uncertainty, $\Delta E$, as $\Delta E/\nu \sim
\hbar$. Hence, the decay rate is essentially determined by the
spectral density profile within a bandwidth $\nu$ around its
energy level. According to Kofman and Kurizki,\cite{kurizki}
Eq.~(\ref{gamma}) constitutes a universal result: frequent
measurements on a given initial state generally lead to its
dephasing through randomization of the corresponding phase.
Depending on whether $\nu$ is much larger or much shorter than the
spectral density width with respect to the center of gravity of
$G(\omega)$, either QZE or AZE is expected, respectively.
In this regard, 1D systems have been largely considered
theoretically,\cite{pascazio1,pascazio2}
while multidimensional realistic
systems are almost unexplored.
Furthermore, Kofman and Kurizki\cite{kurizki2} have already discussed
the idea of using QZE to control the decay rate of quantum systems
coupled to a continuum reservoir by means of weak perturbations.
The experiment carried out by Fischer {\it et al.},\cite{fischer}
with cold atoms initially trapped in an optical lattice, precisely
shows the appearance of QZE and AZE by
repeatedly on/off switching of the coupling between an almost
bound state and a continuum. On the other hand, it has also been
shown\cite{scully} that periodic coherent pulses acting between
the decaying level and an auxiliary one, can either inhibit or
accelerate the decay into a reservoir. In chemical reactivity,
evidences of AZE have been shown by Prezhdo,\cite{prezhdo} who
also considered this mechanism as a novel route to quantum controlling
in chemistry. This is a step towards the
control of decoherence and entanglement in molecular systems which
can handle large amounts of quantum
information.\cite{Viola-Lloyd:98,Vitali-Tombesi:99,kurizki3}

In this work, on the contrary, a mechanism is proposed to quantum
controlling the time-evolution of several competing
fragmentation processes of an isolated molecule in gas phase,
based on the QZE.
Thus, by doing repeated measurements on one of the dissociation channels
the outcome of an unimolecular fragmentation can be controlled.
Isolated molecular systems allow to interpret experimental results
based on realistic multidimensional models, allowing accurate quantum
dynamical studies.

\begin{figure}
 \includegraphics[width=8cm]{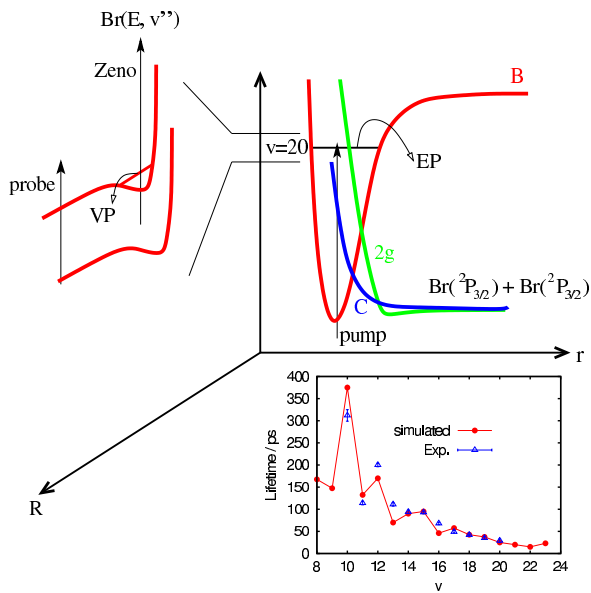}
 \caption{\label{cartoon} Schematics of the VP/EP fragmentation process
  considered here, including the pump/probe and Zeno-type pulses (see
  text for details). In the inset, measured\cite{taylor} and
  simulated\cite{sanz1} total lifetimes.}
\end{figure}

To this end, rare gas-halogen diatom  van der Waals complexes
(Rg$-$X$_2$) are ideal systems displaying a natural separation
between the fast intramolecular X$_2$ vibration and the slow
intramolecular Rg$-$X$_2$ motions.\cite{vdw}
The van der Waals (vdW) complex is
prepared in the $B$ excited electronic and vibrational $v$ state,
Rg$-$X$_2$($B,v$), by the optical $B\leftarrow X$
electronic transition. The vibrational energy, initially deposited
in the X$_2$ subunit, flows towards the weak Rg$-$X$_2$ mode
producing the fragmentation through vibrational predissociation (VP) ,
\begin{eqnarray}\label{VP}
{\rm Rg}-{\rm X}_2(B,v) \rightarrow {\rm Rg} + {\rm X}_2(B, v'< v) .
\end{eqnarray}
The dissociation process is monitored by detecting the
X$_2(B,v')$ fragments by a second probe laser. The presence of
the Rg atom induces electronic couplings in the X$_2$ subunit.
This may eventually produce electronic predissociation (EP) via double
fragmentation,
\begin{eqnarray}\label{EP}
{\rm Rg}-{\rm X}_2(B,v) \rightarrow {\rm Rg} +
{\rm X}(P_{3/2})+ {\rm X}(P_{3/2}) ,
\end{eqnarray}
indicated schematically in Fig.~\ref{cartoon}.
This process was first detected for Ar-I$_2$
complex.\cite{levy1,levy2,zewail,klemperer}
The Ne-Br$_2$ system
constitutes a benchmark since the existing VP/EP competition has
been extensively studied. Numerical results obtained
using 3D potential energy surfaces issued from {\it ab initio}
calculations\cite{sanz1} have been found to be in excellent
agreement with experimental complex lifetimes,\cite{taylor} as
seen in the inset of Fig.~\ref{cartoon}. In this figure,
oscillations are attributed to the EP process due to the
Frank-Condon factors between the electronic $B$ state and the
dissociative 2$_g$ and $C$ states of Br$_2$. As expected from
the energy gap law,\cite{beswick} the VP rate increases
monotonically with $v$; for $v > 15$, it becomes more efficient
and dominates. Based on the excellent realistic model describing
the experimental results,
here we study the NeBr$_2$($B$,$v=$20) decay
using  3D wave packet (WP) simulations under frequent
measurements. As is shown below, this allows us to slow down the
otherwise dominant VP for $v$=20 in favor of the slower EP, which
involves two different dissociative electronic states, the 2$_g$
and $C$ states of Br$_2$.

Let us first only consider the VP process for Ne-Br$_2($B$,v=20)$
in order to illustrate the QZE and AZE.
This is carried out by expanding the WP as\cite{octavio2}
\begin{eqnarray}\label{VP-expansion}
 \vert \Psi (t)\rangle =\sum_{v'} \Phi_{v'}(R,\theta, t)\,
  \varphi_{v'}(r) .
\end{eqnarray}
Here, $\varphi_{v'}(r)$ denote the vibrational eigenfunctions of
the Br$_2$($B$) subunit; $ \Phi_{v'}(R,\theta, t)$ depends on the
distance $R$ between Ne and the Br$_2$ center-of-mass and the
angle $\theta$ between the two Jacobi vectors, ${\bf r}$ and ${\bf
R}$. The predissociation lifetimes are of the order of 10-100
picoseconds, while the quadratic behavior of the decay
occurs at a much shorter
time scale (of the order of femtoseconds or shorter). The
inclusion of a new pulse (Zeno pulse in Fig.~\ref{cartoon})
in the propagation is computationally very
demanding, and a first order perturbation approach is adopted for
simplicity to deal with the corresponding absorption.
 Within this approximation, the effect of the Zeno
pulse can be described by two limiting cases. First, the
population depletion method, where
 the pulse fully promotes the
dissociative $\Phi_{v-1}(R,\theta, t)$ component of the WP to a
higher electronic state by optical pumping. This is simulated by
projecting the WP on the different vibrational states and then
removing the $v-1$ component at each measure time $n\tau$. Second, the dephasing or
randomization method where the measurement introduces a random
phase in the $\Phi_{v-1}(R,\theta,t)$ component as a perturbation,
simulated by multiplying $\Phi_{v-1}(R,\theta,t)$ by a random
phase depending on the coordinates $R$ and $\theta$.

\begin{figure}
 \includegraphics[width=8cm]{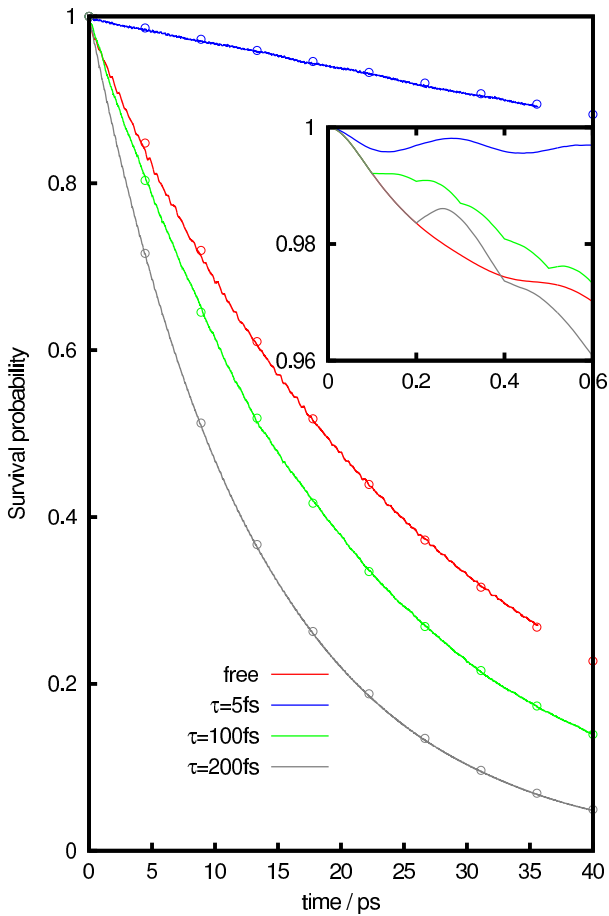}
 \caption{\label{tau-effect} Survival probability as a function of time
  for the free (unperturbed) evolution (red) and using different time
  intervals $\tau$ for the randomization procedure.
  In the inset, enlargement on the short-time dynamical regime.
  Open circles are fits to the exponential decay of Eq.~\ref{pt},
  with $\gamma_0=$ 0.098 (free) and $\gamma(\tau)=$ 0.008 ($\tau$=5 fs),
  0.13 ($\tau$=100 fs) and 0.199 ($\tau$=200 fs),
  all given in cm$^{-1}$.}
\end{figure}

The effect of these two measurement schemes on $P(t)$ can be explained
by formally expanding the WP as
\begin{eqnarray}
\vert \Psi(t)\rangle = a_b(t) e^{-iE_bt/\hbar} \vert b\rangle +
\sum_f a_f(t) e^{-iE_ft/\hbar} \vert f\rangle ,
\end{eqnarray}
where $\vert b \rangle$ and $ \vert f\rangle $ denote the initial bound
state and final dissociative states of $\hat{H }_0$, respectively.
At very short times, the decay of $\vert b\rangle$ into some final
$\vert f\rangle$ is described by first-order perturbation theory
as\cite{cohen}
\begin{eqnarray}\label{first-order-transitions}
{d a_b\over d t} &=& -{i\over \hbar} \sum_f V_{bf}\, a_f\,
e^{-i(E_b-E_f)t/\hbar}.
\end{eqnarray}
By means of a series of repeated measurements at time intervals
$\tau \to 0$, it is possible to freeze the evolution of the
initial state, slowing down the transition by destroying the
coherence on the right-hand side of
Eq.~(\ref{first-order-transitions}) (Ref.~\onlinecite{peres}) (for each
$(R,\theta)$ pair, there is a given positive/negative sign for the
probability amplitude and the integrated amplitude vanishes). As
seen in Fig.~\ref{tau-effect}, the typical QZE and AZE
 behavior is numerically
reproduced for Ne-Br$_2$($B,v$=20) using the randomization model
with different values for $\tau$. With the depopulation method
nearly indistinguishable results are also obtained, demonstrating
the robustness of the approximations made. When
considering repeated measurements with short $\tau$, the lifetime
increases by an order of magnitude, from 27~ps in the free
(unperturbed) case to
about 320~ps in the case of $\tau$=5~fs. A similar deceleration is
also found for $\tau < 50$~fs. This is a clear manifestation of the
QZE. For $\tau=100$~fs, the lifetime is 20~ps, approximately the
same as in the free case, while for $\tau> 100$~fs, VP is enhanced,
yielding a lifetime of 13 fs, according to the AZE.
The exponential decays for
different values of $\tau$ fulfill the predicted behavior given by
Eq.~(\ref{pt}). In Fig.~\ref{tau-effect} the fittings to that
expression are shown with open circles.
The QZE is only found for short values of $\tau$, where the decay
is quadratic. The Zeno  behavior is lost after 0.1-0.2 ps, just
when the curvature of the free population changes.
It should be noticed that
the $v-1$ dissociative component of the WP is measured
instead of the initial $v$ component. However, the usual QZE and
AZE behaviors are found. This is easily understood within the randomization
measurement model. The dissociation in this case is slowed down by
introducing a random phase difference between the bound and
dissociative components of the WP. Such a difference can
be applied either to the initial, $v$, or dissociative, $v-1$,
component, leading to the same results, when only one
fragmentation channel is considered.

\begin{figure}
 \includegraphics[width=7cm]{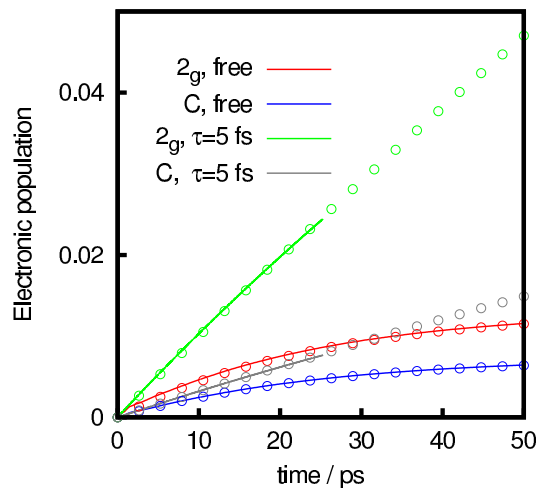}
 \caption{\label{EP-branching} Electronic populations of the 2$_g$
  and $C$ dissociative states of the Br$_2$ as a function of time, due
  to EP for the free dynamics and under repeated measurements with
  $\tau$=5~fs. Solid lines correspond to numerical
  calculations. Open circles correspond to fittings of the form
  $Q^\tau_\alpha(1-e^{-\gamma(\tau) t} )$, with $\gamma_0$=0.106
  cm$^{-1}$, $\gamma(\tau$=5~fs)=0.0083 cm$^{-1}$. The
  branching ratios are $Q^0_{2g}=$0.0133, $Q^0_{C}=$0.0074,
  $Q^5_{2g}=$0.325 and $Q^5_{C}=$0.103.}
\end{figure}

In the case of several fragmentation channels, measuring the
initial state would slow down the total decay towards all the
dissociation channels. On the contrary, measuring one of the
dissociative states, would slow down selectively a given
dissociation channel leaving nearly unchanged the other ones. This
scenario is precisely the one used here when considering the competition
of the VP channel described above together with the EP channel towards
the 2g and C dissociative states of Br$_2$. It is therefore expected
that the slow electronic fragmentation described by Eq.~(\ref{EP}) can
be enhanced by slowing down the dominant VP channel.
To deal with the two dissociation mechanisms, the total WP is written
as\cite{octavio3}
\begin{eqnarray}
 |\Psi(t)\rangle = \sum_\alpha \Phi_\alpha(r,R,\theta,t)\,
  \chi_\alpha,
\end{eqnarray}
where $ \chi_\alpha$ ($\alpha=B$, 2$_g$ and $C$) is the electronic
wave function and $\Phi_\alpha(r,R,\theta)$ is expanded on a grid
of 256$\times$160$\times$40 points for the three Jacobi
coordinates.\cite{sanz1} Due to the large grid involved,
the calculations have been parallelized in the angle
$\theta$.\cite{octavio3}
The randomization model employed above becomes
unstable numerically for long times. For this reason,
the measurements are simulated by the depopulation method, by
projecting the $\Phi_{\alpha=B}(r,R,\theta)$ component onto the
initial state, removing all other possible contributions
describing the VP dynamics at each $\tau$.
The typical QZE oscillations are again observed as
in the previous case. The slowing down effect introduced by the
repeated measurements is again remarkably  reproduced and rather
important. Since the VP process is slowed down, the whole
lifetime increases.

Under such conditions, EP rates become larger or of the same order
than VP ones. Thus, the electronic population for the two
dissociative electronic channels increases. In Fig.~\ref{EP-branching}
the electronic population
of the Br$_2$  2$_g$ and $C$ dissociative states,
is displayed as a function of time. An analytical fitting
to the exponential law is also shown;
the electronic branching ratios are obtained from this fitting.
The branching ratio of the 2$_g$ state changes from 1.3\% in the
free case to 32.5\% with repeated measurements every $\tau$=5~fs;
for the $C$ state, this ratio changes from 0.74\% to 10.3\%. This
implies the branching ratios of these two states increase by
factors 25 and 14, respectively. This is a clear evidence of how
the Zeno pulse can slow down the otherwise dominant VP, producing an
important enhancement of EP, which leads to a total Br+Br+Ne
fragmentation.

For the particular case studied here, experiments could be
performed by simply adding a third pulse, the Zeno-pulse, to the
usual pump-probe experiment.\cite{taylor}
The frequency of this Zeno-pulse
should be tuned into the $B\rightarrow E$ transition maximizing the
absorption intensity.
It is expected that the intensity should be high to promote the population
of the ($B,v-1$) channel as much as possible.  The Zeno-pulses should
be short, of the order of 1~fs in order to allow a measurement
frequency $\nu$ of the order of 0.2~fs$^{-1}$. The relative phase of the pulses
are not expected to play a major role since the (coherent or incoherent)
 dynamics on the upper $E$ state
is not of interest here.  These effects
should hold for $\tau \rightarrow 0$, and it can be envisaged to
use a continuous laser to generate the Zeno pulse, as recently
proposed by Kurizky and co-workers\cite{Kurizki-CW} in a different
context.
In this way, the direct detection of EP products, namely the atomic
Br$(^2P_{3/2})$ fragments would
allow to measure EP rates directly in absence or by diminishing VP.
Thus, the oscillations of EP rates as a function of $v$ associated
with the electronic $B/2g$ or $B/C$ Frank-Condon factors could
then be better interpreted. This could help to unravel the complex
dynamics in other similar systems, such as the prototypical
Ar-I$_2$.\cite{octavio1}
This working scheme will also allow the other way around, that is,
to enhance VP in cases where EP is dominant, such as for $v$=11 in
NeBr$_2$.

In summary, the QZE-based control mechanism is of rather general
applicability, and only requires two competing decay channels. By
acting  on one of them, by slowing down or accelerating it using
the QZE or AZE, respectively, one can obtain a change of the
branching ratios of one or two orders of magnitude. This quantum control
method is specially well adapted for slow decay dynamics, where
other control methods requiring coherence fails, such as the
coherence control method. When including the Zeno-pulse in the
simulations, there may be an attenuation of the processes described,
but the results obtained are expected to be rather robust and will
not change the major physics here proposed.


Support from the Ministerio de Ciencia e Innovaci\'on (Spain) under
Projects Nos. FIS2010-18132, FIS2010-22082, and CSD2009-00038, and by
Comunidad Aut\'onoma de Madrid (CAM) under Grant No.\ S-2009/MAT/1467
are acknowledged. A. S. Sanz  thanks MiCIn for a ``Ram\'on y Cajal''
Grant. The calculations have been performed in the IFF
and CESGA computing centers with special grants.
We would also like to thank Professor G. Kurizki for a critical reading
of the manuscript.


\end{document}